\documentclass[12pt]{article}
\usepackage{mathrsfs,latexsym,psfig,fullpage}

\newtheorem{theorem}{Theorem}
\newtheorem{lemma}{Lemma}

\newcommand{\myps}[4]
{\begin{figure}[htb] \label{#4}
    \centerline {\psfig{figure=#1,width=#2 cm}}
    \centerline {\parbox{12cm}{\caption{#3}}}
  \end{figure}}

\def\lowE{\raisebox{-0.9ex}{$\scriptstyle\sim$}\kern-0.6em E}

\def\bra#1{\left<#1\right|}
\def\ket#1{\left|#1\right>}
\def\scalar#1#2{\left<#1|#2\right>}
\def\dmin{\mathop{\rm d}_{\rm H min}\nolimits}
\def\dH{\mathop{\rm d}_{\rm H}\nolimits}
\def\dHnor{\mathop{\underline{\rm d}}_{\rm H}\nolimits}

\def\trace{\mathop{\rm Tr}\nolimits}

\begin{document}
\title{Performances of Block Codes Attaining the Expurgated and Cutoff Rate 
Lower Bounds on the Error Exponent of Binary Classical-Quantum 
Channels\footnotetext{This paper has been presented at the Fifth International 
Conference on Quantum Communication Measurement \& Computing Capri, 
Italy - July 3-8 2000.}}

\author{Pawel Wocjan,
Dejan E.~Lazic\footnotemark, and Thomas Beth\\
{\small Institut f{\"u}r Algorithmen und Kognitive Systeme,
Universit{\"a}t Karlsruhe}\\ {\small Am Fasanengarten 5,
D--76\,131 Karlsruhe, Germany}\\
\addtocounter{footnote}{-1}\footnotemark\ {\small FZI Forschungszentrum 
Informatik,}\\ {\small Haid-und-Neu-Str.~10-14, D--76\,131 Karlsruhe, Germany}}

\maketitle

\abstract{\emph{
A conceptually simple method for derivation of lower bounds on the error 
exponent of specific families of block codes used on classical-quantum 
channels with arbitrary signal states over a finite Hilbert space is presented.
It is shown that families of binary block codes with appropriately 
rescaled binomial multiplicity enumerators used on binary
classical-quantum channels and decoded by the suboptimal decision rule 
introduced by Holevo attain the expurgated and cutoff rate lower bounds on the 
error exponent.}
}

\section{Introduction}
In information theory the error exponent and its bounds, describing the
exponential behavior of the decoding error probability, are 
important quantitive characteristics of channel performances. 
One can say that ever since Claude Shannon published his famous 
1948 papers \cite{shan}, information theorists mostly used and developed his 
\emph{ensemble averaging} technique of obtaining an asymptotic in code length 
upper bound on the overall probability of block decoding error for the optimal 
channel block code. This central technique of information theory is usually
referred to, not very appropriately, as ``random coding'' \cite{viterbi}.
Formally, it consists of calculating the average 
probability of block decoding error over the ensemble of all sets of code 
words that are possible over the encoding space. This technique was
adopted to \emph{classical-quantum channels} in \cite{cap1a} \cite{cap1b} 
\cite{cap2} \cite{relFunPure} \cite{hol}. Especially, Holevo 
obtained the expurgated and cutoff rate lower bounds on the error
exponent of classical-quantum channels with arbitrary signals states over a
finite Hilbert space. The most unsatisfactory aspect of the ensemble averaging 
technique is that it can not determine the performance of a \emph{specific 
familiy} of codes used on the channel considered, nor can it determine the 
requirements that a familiy of channel block codes should meet in order 
to attain the probability of error guaranteed by the channel error exponent or
its lower bounds.

However, as shown in \cite{laz} for classical channels, it is possible to 
define error exponents and capacity measures for specific code families. In 
this paper we present a direct and conceptually simple method for derivation of
lower bounds on the error exponent of specific families of block codes used on 
classical-quantum channels with arbitrary signal states over a finite 
Hilbert space. The (indirect) ensemble averaging technique uses the notion of an
ensemble of all possible codes, thus concealing the requirements which a
code family should meet, in order to have a positive error exponent and at best 
attain the channel error exponent. These requirements are now stated precisely
by using the \emph{method of multiplicity enumerators}. We show that families 
of binary block codes with appropriately 
\emph{rescaled binomial multiplicity enumerators} attain the expurgated and 
cutoff rate lower bounds on the error exponent of binary classical-quantum 
channels even if decoded by a suboptimal decision rule introduced by Holevo.

The paper is organized as follows. In Section~\ref{CQC} we introduce the 
necessary definitions and notations to describe classical-quantum channels and
give an overview of the results obtained so far by the ensemble averaging 
technique. In Section~\ref{ME} the method of multiplicity enumerators is 
presented. Finally, in Section~\ref{CRB} and Section~\ref{EB} we show that 
families of binary block codes with appropriately rescaled binomial 
multiplicity enumerators attain the expurgated and cutoff rate lower bounds on 
the error exponent of binary classical-quantum channels.

\section{Classical-Quantum Channels}\label{CQC}
Let $\mathscr{C}$ denote a classical-quantum (c-q) channel 
\cite{cq} over a finite input alphabet $\mathscr{A}=\{a_1,\ldots, a_q\}$, 
$|\mathscr{A}|=q \, (<\infty )$, with arbitrary channel output signal states 
$S_1,S_2,\ldots,S_q$, given by density matrices over a finite dimensional 
Hilbert space. For an overview of other possible channel models arising in 
quantum information theory see \cite{ben}.

A code word $\mathbf{x}_m=(x_{m1},x_{m2},\ldots,x_{mN})$,
$x_{mn}\in\mathscr{A}$, $m=1,\ldots,M$, transmitted
over the channel $\mathscr{C}$ induces the product state 
$\mathbf{S}_m = S_{m1} \otimes S_{m2} \otimes \ldots \otimes S_{mN}$
at the channel output. A \emph{block code} $\mathscr{B}$ of length $N$ and size
$M\le q^N$ is a collection $\{{\bf x}_1,{\bf x}_2, \ldots, {\bf x}_M\}$ of $M$ 
code words over $\mathscr{A}$ of length $N$, where $R=\log_2 (M)/N$ is the 
\emph{code rate} of $\mathscr{B}$. The quantum decoder for $\mathscr{B}$ used 
over $\mathscr{C}$ is statistically characterized by a 
\emph{quantum decision rule} 
${\mathscr{D}}=(\mathbf{D}_1,\mathbf{D}_2,\ldots, \mathbf{D}_M)$ which is
a collection of $M$ \emph{decision operators} satisfying 
$\sum_{m=1}^M \mathbf{D}_m \le I$.

Providing these operators a quantum decoder makes the measurement and the
decision simultaneously, in contrast to a classical decoder which makes its
decision on the already measured channel output \cite{kraus} \cite{hel}.

The conditional probability that the quantum decoder makes the decision in 
favor of the code word $\mathbf{x}_j$ when the code word $\mathbf{x}_m$ 
($m\neq j$) is transmitted over $\mathscr{C}$ is given by
\begin{equation}
  e_m({\bf x}_j) =\trace \mathbf{S}_m \mathbf{D}_j\,,
\end{equation}
where $\trace(\cdot)$ denotes the trace of a matrix. The quantity 
$e_m({\bf x}_j)$ is called the \emph{error effect} of the code word 
${\bf x}_j$ on the probability of erroneous decoding, $P_{\rm em}$, of the 
transmitted code word ${\bf x}_m$ \cite{laz}. Consequently the
\emph{word error probability} is given by
\begin{equation}\nonumber
   P_{\rm em} = P[\hat{{\bf x}} \neq {\bf x}_m \mid {\bf x}_m] =
   \sum_{j=1\atop j\neq m}^M e_m({\bf x}_j)\,, \quad m=1,\ldots,M, 
\end{equation}
where $\hat{{\bf x}}$ is the decision made by the rule $\mathscr{D}$.
The \emph{overall block decoding error probability}, $P_{\rm e}$, can be 
expressed in the following form
\begin{equation}\label{obdep}
   P_{\rm e} = \sum_{m=1}^M P[{\bf x}_m] P_{\rm em} =
   {1\over M} \sum_{m=1}^M P_{\rm em} = {1\over M} \sum_{m=1}^M 
   \sum_{j=1\atop j\neq m}^M e_m({\bf x}_j) \,,
\end{equation}
since all a priori code word probabilities $\{P[\mathbf{x}_m]\}_{m=1}^M$ are 
assumed to be equal which is the usual case in channel coding theory.

The design of a communication system for a channel $\mathscr{C}$ consits of the 
choice of the block code $\mathscr{B}$ and the decision rule $\mathscr{D}$. 
Once a code has been choosen, the \emph{optimal decision rule} 
$\mathscr{D}_{\rm opt}$ minimizes $P_{\rm e}$ for this code. For all codes 
used on a memoryless classical channel the optimal decision rule is the
\emph{Maximum-likelihood} decoding. In general, the optimal decision rule on 
classical-quantum channels is not known \cite{hol}. However, conditions for 
the optimal quantum decision rule are given in \cite{hel} \cite{hol73} 
\cite{yuen}.

For a given code length $N$ and fixed code rate $R$ there exists at least one 
\emph{optimal block code} $\mathscr{B}_{\rm opt}$ producing the minimal 
overall block decoding error probability $P_{\rm e opt}$ on the channel 
$\mathscr{C}$, if decoded by the corresponding optimal decision rule 
$\mathscr{D}_{\rm opt}$. 

\subsection{Error Exponent}
Asymptotic reliability performances (for $N\rightarrow \infty$) of the channel 
$\mathscr{C}$ can be expressed by the \emph{reliability function} and the
\emph{channel capacity}. The reliability function, or \emph{error exponent}
\cite{gal}, is 
given by
\begin{equation}\label{errorExp}
   E(R)=\lim_{N\rightarrow\infty} 
         \sup \left\{ -{1\over N} 
                      \log_2 [ P_{\rm e opt}(R,N)] 
   \right\} \,,
\end{equation}
so that 
\begin{equation}\label{Peopt}
  P_{\rm e} \ge P_{\rm e opt}(R,N) = 2^{-N\cdot E(R) + o(N)} \, .
\end{equation}
If the error exponent of the channel $\mathscr{C}$ has positive values, then 
the channel capacity $R_c$ represents the largest code rate at which 
$E(R)$ is still positive. Otherwise the capacity of the 
channel is zero. For all code rates smaller than $R_{\rm c}$ there
exists an infinite sequence of block codes and correponding decoding rules 
such that $P_{\rm e}$ decreases exponentially to zero (according to 
(\ref{Peopt})) with $N\rightarrow\infty$.

The coding theorem determines the asymptotic reliability performances of the
c-q channel and is composed of two parts: the direct part and the converse 
part. The direct part has been established in \cite{cap1a} \cite{cap1b} 
\cite{cap2} by determining the c-q channel capacity as
\begin{equation}\label{cap}
  R_{\rm c} = \max_{\mathbf{P}} \left[
   {\rm H}\!\left(\sum_{i=1}^q p_i S_i\right) - 
   \sum_{i=1}^q p_i {\rm H}(S_i)\right]\,,
\end{equation}
where the maximum is taken with respect to all a priori probability 
distributions $\mathbf{P}=(p_1,p_2,\ldots,p_q)$ on the input alphabet 
$\mathscr{A}=\{a_1,\ldots, a_q\}$, and ${\rm H}(S)=\mbox{Tr}(S\log_2(S))$ 
denotes the von Neumann entropy. This result
has been derived by the \emph{random coding argument} 
\cite{shan} \cite{bla} \cite{cov}
and the concept of \emph{typical subspace} \cite{typSub}. The weak converse of
the channel coding theorem states that the overall block decoding error 
probability does not approach zero when the transmission rate is above the 
channel capacity $R_{\rm c}$. Its proof \cite{BoundOnCapacity} is based on the
combination of the classical Fano's inequality and Holevo's upper bound 
\cite{HolevosBound} on the classical mutual information between the input and 
the output of a c-q channel followed by a quantum measurement.
Holevo's bound can also be derived from the monotonicity 
of the quantum relative entropy \cite{Lindblad} \cite{Uhlmann}. The 
\emph{strong converse} of the coding theorem \cite{strongConverse} 
states that the overall block decoding error probability increases 
exponentially with $N$ up to one for any family of codes with rate $R$ 
above the channel capacity $R_{\rm c}$. 

\subsection{Random Coding Lower Bound}
In \cite{relFunPure} the \emph{random coding exponent} $\lowE_{\rm r}(R)$
\begin{equation}\label{holErrorExp}
  \lowE_{\rm r}(R) = \max_{0 \le s\le 1} \max_{\mathbf{P}} 
  [\mu(s,\mathbf{P}) - s R] \le E(R)
\end{equation}
has been conjectured as a lower bound on the error exponent of the c-q 
channel $\mathscr{C}$, where
\begin{equation}
  \mu(\mathbf{P},s) = -\log_2 
   \trace\left[
     \left(\sum_{i=1}^q p_i S_i^{{1\over 1+s}}\right)^{1+s}\right] \,,
\end{equation}
and the second maximum in (\ref{holErrorExp}) is taken over all a priori 
probability distributions $\mathbf{P}=(p_1,p_2,\ldots,p_q)$ on the input 
alphabet $\mathscr{A}=\{a_1,\ldots, a_q\}$.
The corresponding upper bound on the overall block decoding error 
probability is
\begin{equation}\label{holPeopt}
  P_{\rm e opt}(R,N) \le a\cdot 2^{-N \cdot \lowE_{\rm r}(R)+o(N)}\,,
\end{equation}
where $a$ is a positive constant.

For a \emph{quasi-classical channel}, i.e.~a c-q channel with commuting
signal states $S_i$, the bound (\ref{holErrorExp}) 
is proved with $a=1$ \cite{gal}.
For a \emph{pure state channel}, i.e.~a channel with pure signal states $S_i$, 
the bounds (\ref{holErrorExp}) and (\ref{holPeopt}) hold with
$S_i^{1/(1+s)}=S_i$ (since each $S_i$ is a projector) and $a=2$ 
\cite{relFunPure}. The proof of the random coding bound (\ref{holErrorExp}) 
in full generality, i.e.~for c-q channels with arbitrary 
signal states $S_i$, remains an open question \cite{hol}. However, in this case
the largest code rate $R$ at which $\lowE_{\rm r}(R)$ attains zero is equal to 
channel capacity given by (\ref{cap}) \cite{hol}.

\subsection{Suboptimal Decision Rule}
In \cite{hol} the suboptimal decision rule 
$\mathscr{D}_{\mathcal{H}}$ has been introduced for any block code 
$\mathscr{B}$ used on the c-q channel $\mathscr{C}$ with arbitrary signal 
states $S_i$ where
\begin{equation}\label{DH}
   \mathscr{D}_{\mathcal{H}} =
   (\mathbf{D}_1,\mathbf{D}_2,\ldots,\mathbf{D}_M)\,,\quad
   \mathbf{D}_m =
     \left(\sum_{j=1}^M \mathbf{S}_j^r\right)^{-{1\over 2}}
     \mathbf{S}_m^r \,
     \left(\sum_{j=1}^M \mathbf{S}_j^r\right)^{-{1\over 2}}\,,
\end{equation}
$m=1,2\ldots,M$, and $r$ can take any value from the interval $(0,1]$. 
Furthermore, it has been shown that this decision rule
produces a word error probability $P_{\rm em}$ 
upperbounded by
\begin{equation}\label{subopt}
  P_{\rm em} \le \tilde{P}_{\rm em} =
    \sum_{j=1\atop j\neq m}^M
      {\rm Tr}\, \mathbf{S}_m^{1-r} \mathbf{S}_j^r \,.
\end{equation}

\subsection{Cutoff Rate Lower Bound}
By using the decision rule $\mathscr{D}_{\mathcal{H}}$ given by (\ref{DH}) with
$r=\frac{1}{2}$ and the 
random coding argument the \emph{cutoff rate lower bound} on the error 
exponent of the c-q channel $\mathscr{C}$
\begin{equation}\label{cutlowerbound}
  \lowE_{\rm cut}(R) = \max_{\mathbf{P}} [\mu(1,\mathbf{P})] - R \le E(R)\,,
\end{equation}
has been obtained \cite{hol}.

The code rate $R$ at which $\lowE_{\rm cut}(R)$ attains zero is the 
\emph{quantum cutoff rate} \cite{ban}, the quantum analogy of the classical
cutoff rate \cite{bla}. It is given by
\begin{equation}
  R_0 = \max_{\mathbf{P}} \mu(1,\mathbf{P}) = 
    \max_{\mathbf{P}} \left\{ -\log_2
      \trace\left[\left(\sum_{i=1}^q p_i \sqrt{S_i}\right)^2\right]\right\}
  \,,
\end{equation}
and represents a lower bound on the channel 
capacity, i.e. $R_0 \le R_{\rm c}$ \cite{hol}.

\subsection{Expurgated Lower Bound}
In \cite{hol} the \emph{expurgated error exponent} $\lowE_{ex}(R)$ as a lower 
bound on the error exponent 
\begin{equation}\label{ExpErrExp}
  \lowE_{\rm ex}(R) = \max_{s\ge 1} \max_{\mathbf{P}} 
  [\tilde{\mu}(\mathbf{P},s) - s R] \le E(R)\,,
\end{equation}
has been obtained by expurgating poor code words from the codes in the 
ensemble \cite{gal}. The function $\tilde{\mu}(\mathbf{P},s)$ in 
(\ref{ExpErrExp}) is given by 
\begin{equation}
  \tilde{\mu}(\mathbf{P},s) = -s\log_2\left[
    \sum_{i,j=1}^q p_i p_j 
    \left( \trace \sqrt{S_{i_{}}} \sqrt{S_{j_{}}} \right)^{{1\over s}}
  \right]\,.
\end{equation}
The corresponding upper bound on the overall block decoding probability is
given by
\begin{equation}
  P_{\rm e opt}(R,N) \le 4\cdot 2^{-N \cdot \lowE_{\rm ex}(R)+o(N)} \,.
\end{equation}

\subsection{Binary Classical-Quantum Channels}
For binary c-q channels $\mathscr{C}$, i.e.~channels over the input alphabet 
$\mathscr{A}=\{0,1\}$ with the corresponding signal states 
$S_1$ and $S_2$, the expurgated and cutoff rate lower bounds can be computed
explicitly. By introducing the channel parameter 
$c=\trace \sqrt{S_1} \sqrt{S_2}$ of the binary c-q channel 
$\mathscr{C}$ the \emph{cutoff rate lower bound} (\ref{cutlowerbound}) can be
expressed as
\begin{equation}
  \lowE_{\rm cut}(R) = \max_{\mathbf{P}}
    [-\log_2 (p^2 + (1-p)^2 + 2 p (1-p) c)-R] \,,
\end{equation}
and the \emph{expurgated lower bound} (\ref{ExpErrExp}) as 
\begin{equation}\label{binaryExpErrExp}
  \lowE_{\rm ex}(R) = \max_{s\ge 1} \max_{\mathbf{P}}
    [-s\log_2 (p^2 + (1-p)^2 + 2 p (1-p) c^{\frac{1}{s}})-s R] \,.
\end{equation} 
It is easily seen that in the binary case the cutoff rate lower
bound and the expurgated lower bound are attained for the probability 
distribution $\mathbf{P}=(p={1\over 2}, 1-p={1\over 2})$. 
Consequently, the cutoff rate lower bound reduces to
\begin{equation}\label{binaryCutOffRateLowerBound}
  \lowE_{\rm cut}(R) = R_0 - R
\end{equation}
where $R_0$ is the cutoff rate given by
\begin{equation}\label{binaryCutOffRate}
  R_0 = 1-\log_2(1+c)\,.
\end{equation}
The expurgated lower bound is given by
\begin{equation}\label{implicit}
  \lowE_{\rm ex}(R) = \tilde{\mu}(\tilde{s}_R)-\tilde{s}_R R\,,
\end{equation}
where $\tilde{\mu}(s)=\tilde{\mu}(\mathbf{P}=(\frac{1}{2},\frac{1}{2}),s)=-s\,
\log_2\left({1+c^{1/s}\over 2}\right)$ and 
$\tilde{s}_R$ is implicitly given by 
${\partial\over\partial s}\tilde{\mu}(\tilde{s}_R)=R$. 
For rates below the \emph{expurgated rate} $R_{\rm ex}$ given by 
\begin{equation}
  R_{\rm ex}= {\partial\over\partial s}\tilde{\mu}(1) = 
  \tilde{\mu}(1)+{c\over {1+c}} \log_2 c\,,
\end{equation}
the expurgated lower bound is higher than the cutoff rate lower bound. For
rates above the expurgated rate $R_{\rm ex}$ the expurgated and cutoff rate
lower bounds coincide. These results have been obtained in 
\cite{CodingTheoremsQC} for the binary pure state channel. In this case the 
signal states are projectors $S_i=\ket{\Psi_i}\!\bra{\Psi_i}$, for $i=1,2$, and
the channel parameter is $c=\epsilon^2$ where 
$\epsilon=|\scalar{\Psi_1}{\Psi_2}|$.

\section{Multiplicity Enumerators}\label{ME}
In the following a binary c-q channel $\mathscr{C}$, i.e.~a channel
over the input alphabet $\mathscr{A}=\{0,1\}$ with the corresponding signal 
states $S_1$ and $S_2$, will be considered. Over this channel 
binary codes $[N,R]$ of length $N$ and code rate $R=\log_2 (M)/N$ 
($0\le R\le 1$) can be transmitted only.
\begin{lemma}\label{subDec}
  The word error probability $P_{\rm em}$ of the
  code $[N,R]$ used on the channel $\mathscr{C}$ and decoded by the suboptimal
  decision rule (\ref{DH}) with $r=\frac{1}{2}$ is upperbounded by
\begin{equation}
   P_{\rm em} \le \sum_{j=1\atop j\neq m}^M 
     2^{-N [-\dHnor(\mathbf{x}_m,\mathbf{x}_j) \log_2 (c)]}\,,
\end{equation}
where $c=\trace \sqrt{S_1} \sqrt{S_2}$ and 
$\dHnor(\mathbf{x}_m,\mathbf{x}_j)=\dH(\mathbf{x}_m,\mathbf{x}_j)/N$ is the 
normalized Hamming distance between the code words $\mathbf{x}_m$ and 
$\mathbf{x}_j$.
\end{lemma}
Proof: According to the decision rule ${\mathscr D}_{\mathcal H}$ given by
(\ref{DH}) with $r=\frac{1}{2}$, for each pair of code words 
$\mathbf{x}_m$ and $\mathbf{x}_j$ ($j\neq m$) in (\ref{subopt}) holds
\begin{eqnarray}
  \trace \sqrt{\mathbf{S}_m} \sqrt{\mathbf{S}_j} 
  & = & 
  \prod_{n=1}^N \trace \sqrt{S_{mn}} \sqrt{S_{jn}}\\
  & = & 
  \prod_{n=1\atop x_{mn}\neq x_{jn}}^N 
    \trace \sqrt{S_{mn}} \sqrt{S_{jn}}\\
  & = &
  c^{\dH(\mathbf{x}_m,\mathbf{x}_j)} \\
  & = &
  2^{\dH(\mathbf{x}_m,\mathbf{x}_j) \log_2 (c)} \\
  & = &
  2^{-N [-\dHnor(\mathbf{x}_m,\mathbf{x}_j) \log_2 (c)]} \, .
\end{eqnarray}
By summing over all $j\neq m$ the proof is completed. \hfill $\Box$ 

\noindent Consequently, for the suboptimal decision rule 
${\mathscr D}_{\mathcal H}$ given by (\ref{DH}) with $r=\frac{1}{2}$ the 
overall block decoding error probability is upperbounded by the 
\emph{union bound} given by
\begin{equation}\label{uppBound}
   P_{\rm e} \le \tilde{P}_{\rm e} = 
   \frac{1}{M} \sum_{m=1}^M \sum_{j=1\atop j\neq m}^M 
           2^{-N [-\dHnor(\mathbf{x}_m,\mathbf{x}_j)\log_2 (c)]}\,. 
\end{equation}
On the classical binary symmetric channel (BSC) the union bound is also given by
(\ref{uppBound}) but with $c=\sqrt{4 p(1-p)}$, where $p$ is the crossover 
probability.

The presented union bound on the overall block decoding error probability of a 
binary block code $[N,R]$ used on the c-q channel $\mathscr{C}$ and decoded by 
the suboptimal decision rule $\mathscr{D}_{\mathcal{H}}$ with $r=\frac{1}{2}$ 
depends on Hamming distances among the code words and the channel 
parameter $c$. This permits us to use a new method, derived from \cite{laz}, 
in order to estimate the error exponent and the capacity performance of  
specific code families used on the c-q channel $\mathscr{C}$. For this purpose 
some necessary definitions and terms will be introduced.

Every code word $\mathbf{x}_m$ of a binary block code (linear or non-linear)
$[N,R]$ has the Hamming \emph{multiplicity enumerators} 
$\mathbf{M}_{\rm H}(\mathbf{x}_m)$ given by the $(N+1)$-tupel 
\begin{equation}
  \mathbf{M}_{\rm H}(\mathbf{x}_m) = (M_{m0},M_{m1},\ldots,M_{mN})\,,\quad
  \sum_{n=0}^N M_{mn} = M\,,
\end{equation}
where $M_{mn}$ represents the \emph{multiplicity} (number) of code words on the
Hamming distance $n$ from the code word $\mathbf{x}_m$. The multiplicity 
$M_{m0}$ is by definition equal to one.

The \emph{average (expected) multiplicity enumerators} 
$\underline{\mathbf{M}}_{\rm H}[N,R]$ of the block code $[N,R]$ 
is given by 
\begin{equation}\label{averMult}
  \underline{\mathbf{M}}_{\rm H}[N,R] = (\underline{M}_0,\underline{M}_1,
    \ldots,\underline{M}_N)\,,\quad  \sum_{n=0}^N \underline{M}_n = M\,,
\end{equation}
where 
\begin{equation}
  \underline{M}_n = \mbox{E}[\underline{M}_{mn}] = 
                    {1\over M}\sum_{m=1}^M M_{mn}
\end{equation}
represents the average (expected) multiplicity of code words on the Hamming
distance $n$.

The \emph{weight enumerators} $\mathbf{W}_{\rm H}[N,R]$ of the block code 
$[N,R]$ is the $(N+1)$-tuple 
\begin{equation}
  \mathbf{W}_{\rm H}[N,R]=(A_0,A_1,\ldots,A_N)\,,\quad \sum_{n=0}^N A_n = M\,,
\end{equation}
where $A_n$ is the number of code words in $[N,R]$ of the Hamming weight 
equal to $n$. $A_0$ and $A_N$ can take the values $0$ or $1$ only.

For linear codes the multiplicity enumerators of all code words are
equal and coincide with the corresponding weight enumerators. For the 
non-redundant linear binary code of rate $R=1$ ($M=2^N$) weight enumerators 
(and multiplicity enumerators) are given by the corresponding binomial 
coefficients ${N\choose n}$, $n=0,1,\ldots, N$. Since 
\begin{equation}
  \sum_{n=0}^N {N\choose n} = 2^N\,,
\end{equation}
it is obvious that there is no redundant binary block code $[N,R]$ ($R<1$)
with average multiplicity enumerators or weight enumerators given by all 
binomial coefficients ${N\choose n}$, $n=0,1,\ldots, N$. However, 
appropriately \emph{rescaled binomial coefficients} $b{N\choose n}$, where 
$b=2^{N(R-1)}$, satisfy the necessary condition
\begin{equation}
  \sum_{n=0}^N b {N\choose n} = 2^{R N} = M\,,
\end{equation}
for average multiplicity enumerators of redundant block codes.

For an infinite family of block codes \emph{asymptotic performances} can be
considered if the code length $N$ of its members can tend to infinity. In
general, this is
the case for \emph{fixed rate sequences}
\begin{equation}\label{FRS}
  {\rm FRS}(R)=\{[N_1,R],[N_2,R],\ldots,[N_i,R],\ldots \}
\end{equation}
of block codes with the same code rate $R=\log_2(M_i)/N_i$ 
($0<R<1$) and increasing code length, i.e.~$N_i<N_{i+1}$ for $i=1,2,\ldots$. 

In the asymptotic analysis of fixed rate sequences it is convenient to express 
the average multiplicities of the block code $[N_i,R]\in{\rm FSR}(R)$ by its
\emph{average multiplicity exponents} $\mathscr{M}_{\rm H}[N_i,R]$ given by 
the $(N_i+1)$-tupel
\begin{equation}\label{AME}
  \mathscr{M}_{\rm H}[N_i,R]=(\underline{\mathscr{M}}_0^{(i)},
                        \underline{\mathscr{M}}_1^{(i)},\ldots,
                        \underline{\mathscr{M}}_{N_i}^{(i)})\,,\quad
  \sum_{n=0}^{N_i} 2^{N_i \underline{\mathscr{M}}_n^{(i)}}=M_i\,,
\end{equation}
where 
\begin{equation}
  \underline{\mathscr{M}}_n^{(i)} =
  {1\over N_i}\log_2 (\underline{M}_n^{(i)}) \,,\quad n=0,1,\ldots, N_i
\end{equation}
and $\underline{M}_n^{(i)}$ are given by (\ref{averMult}).
For $N_i\rightarrow\infty$ the corresponding asymptotic values (if they exist) 
form an infinite sequence representing the 
\emph{asymptotic average multiplicity exponents} (AAME), 
$\mathscr{M}[R]_{\rm FRS}$, of the FSR($R$)
\begin{equation}\label{AAME}
  \mathscr{M}[R]_{\rm FRS} =
  (\underline{\mathscr{M}}_0, \underline{\mathscr{M}}_1, \ldots,
   \underline{\mathscr{M}}_n, \ldots )
\end{equation} 
where
\begin{equation}
  \underline{\mathscr{M}}_n = 
  \lim_{N_i\rightarrow\infty} {1\over N_i} 
  \log_2 (\underline{M}_n^{(i)})\,,\quad n=0,1,2,\ldots .
\end{equation}
All possible values of the normalized Hamming distance  
$\dHnor = {n\over N_i}$ are in the interval $[0,1]$ and in the 
asymptotic case the values
\begin{equation}\label{dense}
  \underline{\delta}_{\rm H} = \lim_{N_i\rightarrow\infty}
  \left({n\over N_i}\right)
\end{equation}
are dense enough. Consequently, the AAMEs can be replaced by a 
continous function, $\mathscr{M}(\underline{\delta}_{\rm H},R)_{\rm FRS}$, 
which interpolates the discrete values of ${\mathscr{M}_{\rm H}[R]}_{\rm FRS}$.
We will call this continous function of the continous argument 
$\underline{\delta}_{\rm H}$ and the parameter $R$ the 
\emph{interpolated asymptotic average multiplicity exponent} (IAAME) of the
fixed rate sequence FRS($R$) of block codes.
\begin{lemma}
If the average multiplicity enumerators of the codes in the fixed rate
sequence ${\rm FRS}(R)$ are given by
rescaled binomial coefficients
\begin{equation}\label{aBinom}
  \left\{ \underline{M}_n^{(i)} = {a_i(n)\over 2^{N_i (1-R)}}  
  {N_i\choose n}\right\}_{n=0}^{N_i}\,,\quad 
  \sum_{n=0}^{N_i} \underline{M}_n^{(i)} = M_i
\end{equation}
where $a_i(n)=o(2^n)$, than the corresponding IAAME function is given by
\begin{equation}\label{explicitIAAME}
  \mathscr{M}(\underline{\delta}_{\rm H},R)_{\rm FRS}=
    {\rm H}_2(\underline{\delta}_{\rm H}) - (1 - R) \,,
\end{equation}
where ${\rm H}_2(x)=-x\log_2(x) - (1-x)\log_2 (1-x)$ represents the binary 
entropy function.
\end{lemma}
Proof: By replacing (\ref{aBinom}) in (\ref{AAME}) directly 
follows
\begin{equation}\label{hilfslimit}
  \lim_{N_i\rightarrow\infty} {1\over N_i}\log_2 \left[ 
    \frac{a_i(n)}{2^{N_i (1-R)}}{N_i\choose n} \right] \,,\quad  
    n=0,1,\ldots, N_i\,.
\end{equation}
Using the well known asymptotic relation (obtained from the 
Stirling approximation \cite{ham}) 
\begin{equation}
  \lim_{N_i\rightarrow\infty} {N_i\choose n} =
  2^{N_i {\rm H}_2({n\over N_i})-o(N_i)}\,,
\end{equation}
the limit (\ref{hilfslimit}) reduces to
\begin{equation}\label{limit}
  \lim_{N_i\rightarrow\infty} {\log_2(a_i(n))\over N_i} + 
  {\rm H}_2(\underline{\delta}_{\rm H}) - (1 - R)\,.
\end{equation}
Taking into account the conditions on $a_i(n)$ and that $n\le N_i$ the limit in
(\ref{limit}) tends to zero and the proof is completed. \hfill $\Box$

Fixed rate sequences whose codes have average multiplicity enumerators are 
given by (\ref{aBinom}) will be called fixed rate sequences with 
\emph{rescaled binomial average multiplicity enumerators}.

Consider an $[N,R]$ linear systematic binary block code with generator matrix
$G=[I\,|\,A]$, where $I$ is the $K\times K$ identity matrix ($K=R N$) and $A$ 
is a $K\times (N-K)$ matrix of $0$'s and $1$'s. Suppose each entry of $A$ is 
chosen at random, to be $0$ or $1$ with probability $\frac{1}{2}$, and then 
choose one of the codewords ${\mathbf x}$ at random. It is 
known \cite{sloane} (Problems, pp.~287) that
\begin{equation}\label{rlbc}
\begin{array}{l}
  \mbox{Prob}\{{\rm w}_{\rm H}({\mathbf x})=0\}=2^{-K}, \\ 
  \mbox{Prob}\{{\rm w}_{\rm H}({\mathbf x})=n\}=2^{-N}\left\{
    {N\choose n}-{N-K\choose n}\right\},\, 0<n\le N\,.
\end{array}
\end{equation}
The expected multiplicity enumerators of this random linear block code
$[N,R]$ can be obtained by multiplying (\ref{rlbc}) with $M=2^K=2^{R N}$ for
all codewords. Consequently, the IAAME function of a FSR($R$) of these random 
linear block codes is given by (\ref{explicitIAAME}).

Nonrandom linear binary block codes of finite length with average multiplicity 
enumerators closely fitting the rescaled binomial multiplicity enumerators have
been found for $N\le 200$ in \cite{bsc} \cite{kalouti}.

\subsection{Performances of binary block codes attaining the cutoff rate lower 
bound on the error exponent of a binary c-q channel}\label{CRB}
\begin{theorem}[Cutoff rate lower bound]
Fixed rate sequences with rescaled binomial average multiplicity enumerators 
used on the binary c-q channel $\mathscr{C}$ and decoded by the suboptimal 
rule (\ref{DH}) with $r=\frac{1}{2}$ have an error exponent whose lower bound 
is the cutoff rate bound given by (\ref{binaryCutOffRateLowerBound}).
\end{theorem}
Proof: Since all summands
$2^{-N [-\dHnor(\mathbf{x}_m,\mathbf{x}_j)\log_2 (c)]}$ in 
(\ref{uppBound}) are equal for the same value of the normalized 
Hamming distance $\dHnor(\mathbf{x}_m,\mathbf{x}_j)$, the overall 
block decoding error probability (\ref{obdep}) can be upperbounded by 
$$
  P_{\rm e} \le \tilde{P}_{\rm e} = 
  \sum_{n=0}^N \underline{M}_n 2^{-N[-{n\over N}\log_2(c)]}\,,
$$
where $\underline{M}_n$, $n=0,1,\ldots,N$, are the average multiplicities
enumerators (\ref{averMult}) of the code and ${n\over N}$, $n=0,1,\ldots,N$ 
represent all possible values of the normalized Hamming distance $\dHnor$. For
codes from a fixed rate sequence (\ref{FRS}) the average multiplicity exponents
(\ref{AME}) can be used, so that
\begin{eqnarray}\label{uppBoundDD}
  P_{\rm e}^{(i)} \le \tilde{P}_{\rm e}^{(i)}  & = & 
  \sum_{n=0}^{N_i} 
    2^{N_i \underline{\mathscr{M}}_n^{(i)}} 
    2^{-N_i [{n\over N_i}\log_2\left({1\over c}\right)]}\nonumber \\
  & = &
  \sum_{n=0}^{N_i} 
    2^{-N_i [{n\over N_i}\log_2\left({1\over c}\right)-
             \underline{\mathscr{M}}_n^{(i)}]}
  \,,\quad i=0,1,2,\ldots\,.
\end{eqnarray}  
It is obvious that all the exponents in (\ref{uppBoundDD}) should 
be negative for all permissible values of $n$ and $c$, since otherwise the 
bound (\ref{uppBoundDD}) becomes trivial, i.e.
$P_{\rm e}^{(i)}\le A$ with $A\ge 1$. Furthermore, if all exponents are 
negative, than the exponent with the minimal absolute value determines the upper
bound if
$N_i\rightarrow\infty$, so that
$$ \tilde{P}_e^{(i)} = 
   2^{-N_i \min_{0\le n\le N_i} 
    [{n\over N_i}\log_2\left({1\over c}\right)-\underline{\mathscr{M}}_n^{(i)}]
      +o(N_i)} \ge P_{\rm e}^{(i)}\,. $$
By replacing the normalized Hamming distance ${n\over N_i}$ by 
$\underline{\delta}_{\rm H}$ as in (\ref{dense}) and the AME
$\underline{\mathscr{M}}_n^{(i)}$ by the IAAME
$\underline{\mathscr{M}}(\underline{\delta}_{\rm H}, R)_{\rm FSR}$ given by
(\ref{explicitIAAME}), and comparing with (\ref{errorExp}) directly follows
\begin{equation}
  \lowE_{\rm cut}(R) = 
  \min_{0\le \underline{\delta}_{\rm H} \le 1}
    [\underline{\delta}_{\rm H}\log_2\left({1\over c}\right) - 
    {\rm H}_2(\underline{\delta}_{\rm H}) + 1 - R] \le E(R) \,.
\end{equation}
The function 
$\underline{\delta}_{\rm H}\log_2\left({1\over c}\right) - 
{\rm H}_2(\underline{\delta}_{\rm H}) + 1 - R$ has the minimum at 
\begin{equation}\label{effectiveDistance}
  \underline{\delta}_{\rm H eff}={c\over 1+c}\,.
\end{equation}
This minimum is 
\begin{equation}\label{mincutoffrate}
   \lowE_{\rm cut}(R)=1-\log_2(1+c) - R\,,
\end{equation}
so that the overall block decoding error probability decreases exponentially 
to zero for any code rate $R$ below the cutoff rate 
$R_0$ given by (\ref{binaryCutOffRate}). {} \hfill $\Box$

\myps{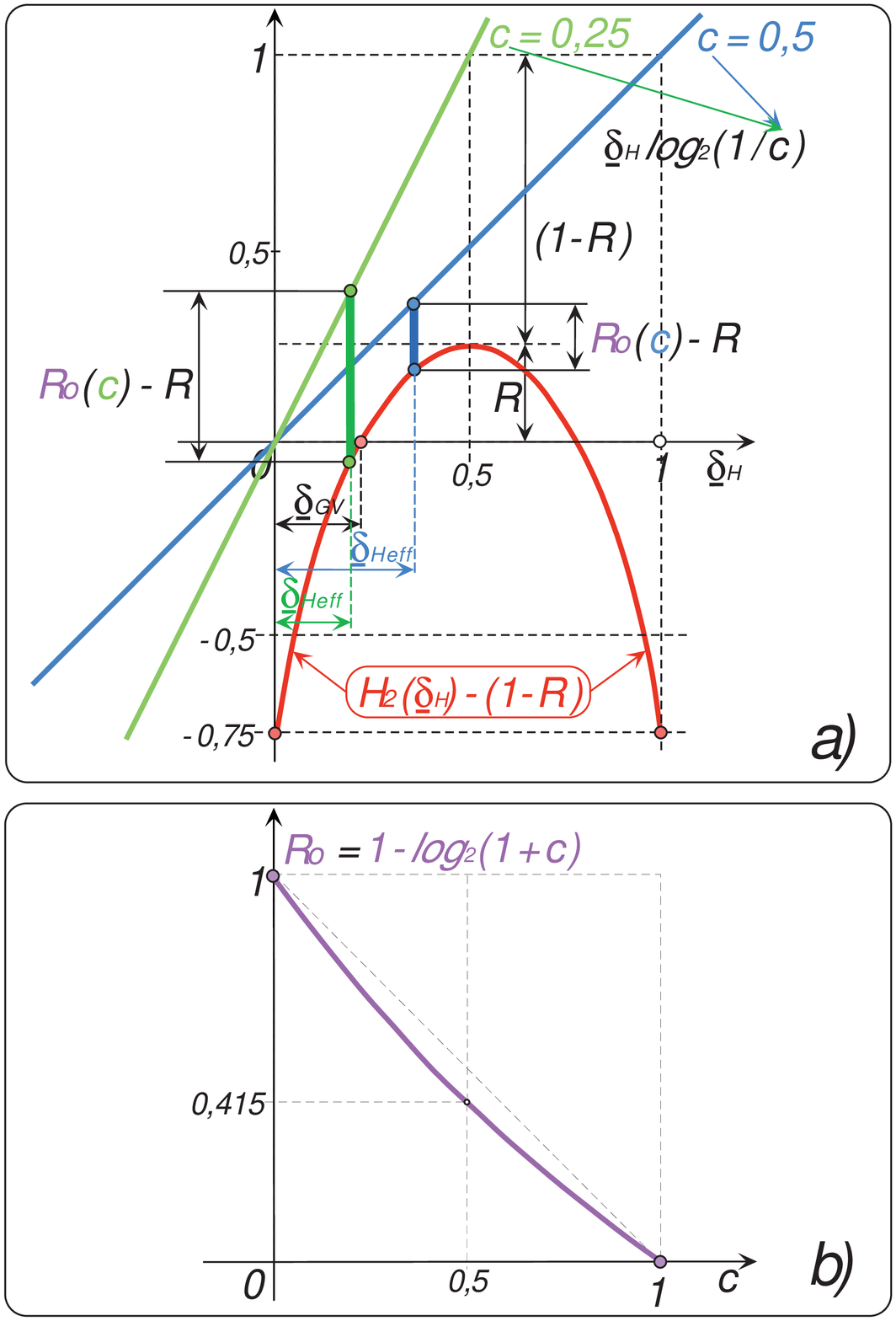}{10}{a) Graphical interpretation of the cutoff rate lower 
bound b) Cutoff rate in dependence on the channel parameter $c$}{CutoffrateFig}
\noindent
Figure~1a gives the graphical interpretation of the method 
of multiplicity enumerators for obtaining the cutoff rate lower bound 
$\lowE(R)_{\rm cut}$ for fixed rate sequences of block codes 
$[N_i,R]$ with rescaled binomial multiplicity enumerators given by 
(\ref{aBinom}). The FSR($R$) is characterized by its IAAME 
${\rm H}_2(\underline{\delta}_{\rm H}) - (1-R)$. The suboptimal decoding rule 
(\ref{DH}) is characterized by the straight line passing through the origin 
with the slope $\log_2\left({1\over c}\right)$. The better the channel (i.e. 
the smaller $c$) the steeper is the slope of the straight line. This is 
illustrated by two values $c=0.25$ and $c=0.5$. The cutoff rate lower bound on 
the error exponent $\lowE_{\rm cut}(R)$ corresponds to the minimal vertical
difference between the straight line characterizing the suboptimal decoding rule
and the IAAME. The minimal difference is attained on the 
\emph{effective Hamming distance} given by (\ref{effectiveDistance}). 
In Figure~1b the dependence of the cutoff rate $R_0(c)$ on 
the channel parameter $c$ is presented.

\myps{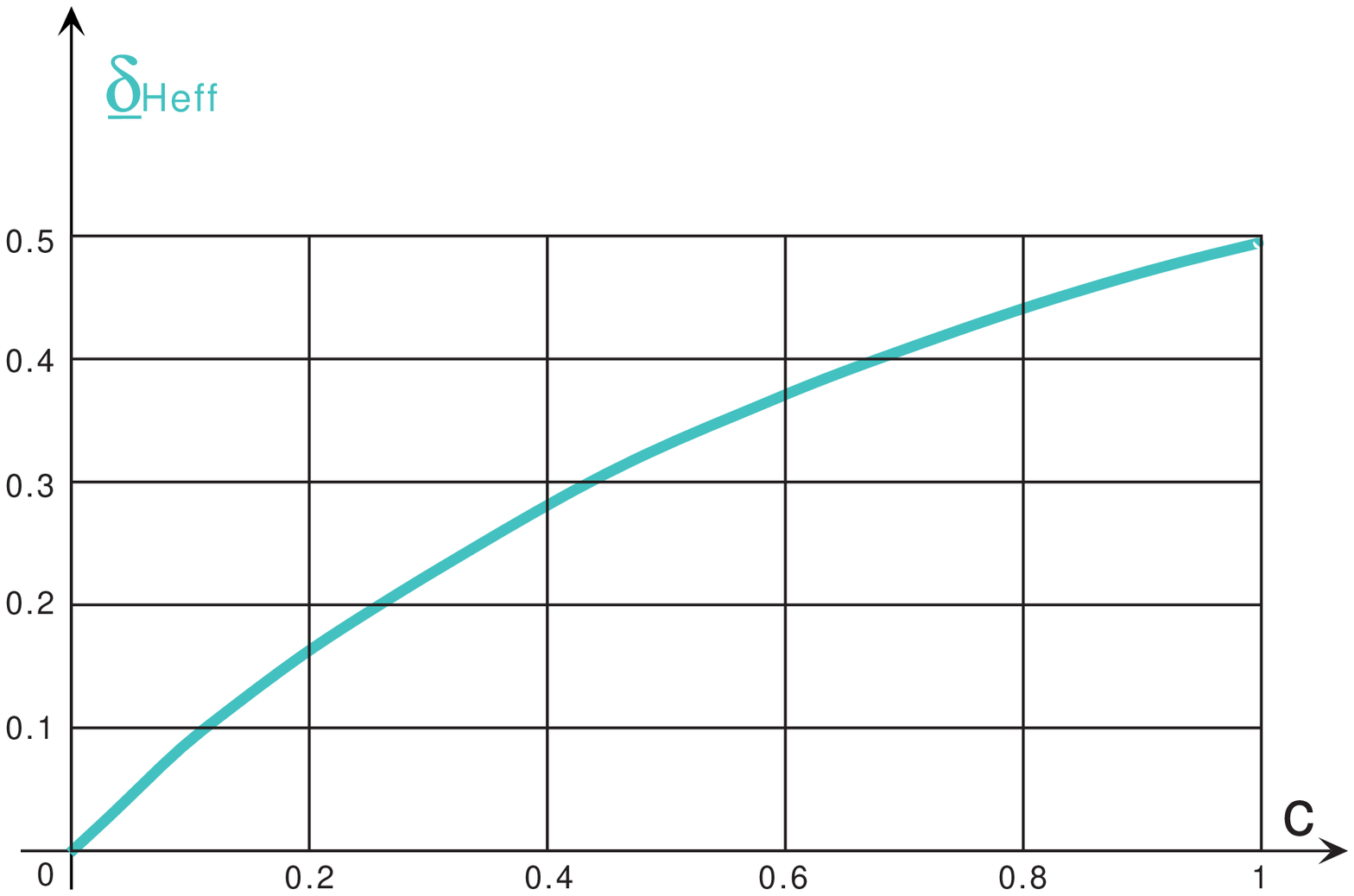}{10}{Effective distance in dependence on the channel parameter
$c$}{EffDistFig}
\noindent
A very important consequence of Theorem~1 is that \emph{only} code words on the 
effective Hamming distance (\ref{effectiveDistance}) determine the 
cutoff rate lower bound. Note that this distance does not depend
on the code rate. Generally, the effective Hamming distance is different 
from the asymptotic \emph{minimal Hamming distance} of the fixed rate sequence 
FSR($R$) given by
\begin{equation}\label{MinNormDist}
  \underline{\delta}_{\rm min} = \lim_{N_i\rightarrow\infty} 
    {\dmin^{(i)}\over N_i}\,,
\end{equation}
where 
\begin{equation}
  \dmin^{(i)}=\min \{\dH(\mathbf{u},\mathbf{v})\mid\mathbf{u},
  \mathbf{v}\in [N_i,R], \mathbf{u}\neq\mathbf{v}\}\,.
\end{equation}
In Figure~2 the effective Hamming distance 
$\underline{\delta}_{\rm H eff}$ in dependence on the channel parameter $c$
is presented.
The better the channel, the smaller is the effective Hamming distance 
$\underline{\delta}_{\rm H eff}$. The worser the channel, the bigger is the
effective distance and the minimal distance does not determine the cutoff rate
lower bound if 
$\underline{\delta}_{\rm min}< \underline{\delta}_{\rm H eff}$.

It follows directly from (\ref{mincutoffrate}) that a fixed rate 
sequence with rescaled binomial multiplicity enumerators used on the 
binary c-q channel and decoded by the suboptimal rule 
$\mathscr{D}_{\mathcal{H}}$, given by (\ref{DH}), with $r=\frac{1}{2}$ has a 
positive cutoff rate lower bound $\lowE_{\rm cut}(R)$ if the channel parameter
$c$ satisfies
\begin{equation}
  c\le 2^{(1-R)}-1\;.
\end{equation}

\subsection{Performances of binary block codes attaining the expurgated lower 
bound on the error exponent of a binary c-q channel}\label{EB}
The \emph{Gilbert-Varshamov lower bound} \cite{sloane} on the asymptotic 
minimal normalized distance (\ref{MinNormDist}) of binary block codes $[N,R]$ 
is given by
\begin{equation}\label{GVB}
  \underline{\delta}_{\rm GV}(R) = {\rm H}_2^{-1}(1-R)\,,
\end{equation}
where ${\rm H}_2^{-1}$ denotes the inverse of the binary entropy 
function in the interval $[0,{1\over 2}]$. Thus, there exist
fixed rate sequences FSR($R$) such that the multiplicities of all code words
$\mathbf{x}_m$, $m=1,2,\ldots,M^{(i)}$, satisfy
\begin{equation}
  M^{(i)}_{mn}=0\,,
\end{equation}
for all Hamming distances $n$ with $n\ge 1$ and 
${n\over N_i}\le\underline{\delta}_{\rm GV}(R)$. Note that all fixed rate 
sequence with IAAMEs which have parts with negative values can be expurgated, 
i.e.~it is possible to remove all code words which correspond to the negative
parts of the IAAME without changing the code rate of the sequence in the 
asymptotic case ($N\rightarrow\infty$). Consequently, the IAAME of a FSR($R$) 
with rescaled binomial multiplicity enumerators satisfying the 
Gilbert-Varshamov bound is given by the \emph{expurgated IAAME}
\begin{equation}\label{zeroBelowGV}
  \mathscr{M}(\underline{\delta}_{\rm H},R)_{\rm FSR} = \left\{ 
  \begin{array}{cl}
    -\infty & \mbox{for } \underline{\delta}_{\rm H} \le 
    \underline{\delta}_{\rm GV}\,, \\
    {\rm H}_2(\underline{\delta}_{\rm H}) - (1 - R) & \mbox{otherwise}.
   \end{array} \right.
\end{equation}
\begin{lemma}
Fixed rate sequences of linear block codes with rescaled binomial multiplicity
enumerators (\ref{aBinom}) satisfy the Gilbert-Varshamov bound (\ref{GVB}).
\end{lemma}
Proof: The average multiplicites of linear codes coincide with the 
corresponding weight enumerators. The weight enumerators are 
natural numbers and cannot take values from the open intervall $(0,1)$. 
Therefore the multiplicity exponents cannot be negative for linear codes.
Consequently, fixed rate sequences of linear binary block codes with 
rescaled binomial average multiplicity enumerators (\ref{aBinom}) automatically
satisfy the expurgated IAAME (\ref{zeroBelowGV}).
{} \hfill $\Box$
\myps{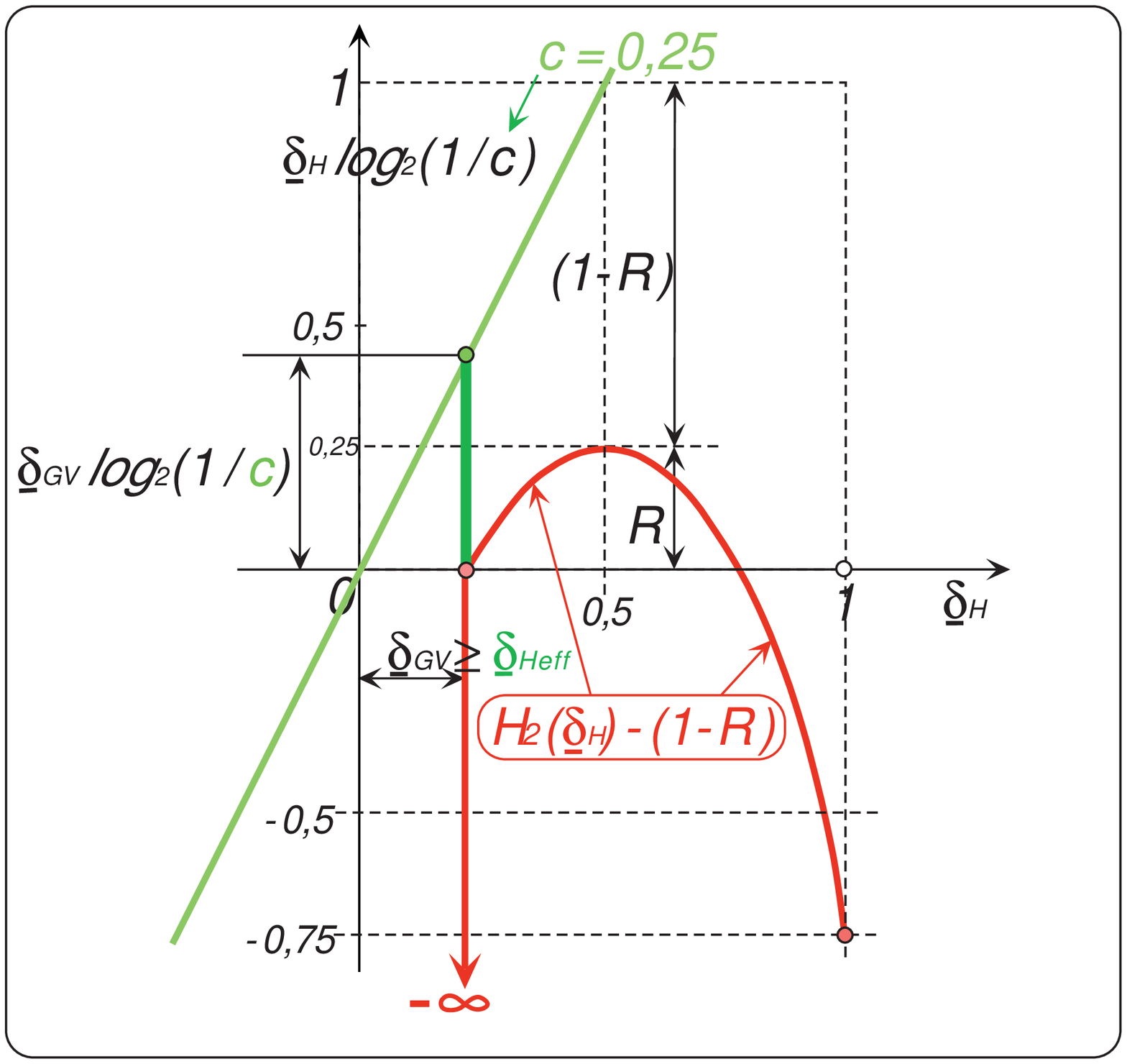}{10}{Graphical interpretation of the expurgated lower 
bound}{ExpErrExpFig}
\begin{theorem}[Expurgated lower bound]
Fixed rate sequences with rescaled binomial multiplicity enumerators 
(\ref{aBinom}) satisfying the Gilbert-Varshamov bound (\ref{GVB}) used on 
the binary c-q channel $\mathscr{C}$ and decoded by the suboptimal 
rule $\mathscr{D}_{\mathcal{H}}$, given by (\ref{DH}), with $r=\frac{1}{2}$ 
have an error exponent whose lower bound is the expurgated bound given by
\begin{equation}\label{GVBH}
  \lowE_{\rm ex}(R)=\log_2\left({1\over c}\right) \cdot 
  \underline{\delta}_{\rm GV}(R)
\end{equation}
for rates $R$ below the expurgated rate $R_{\rm ex}$
\begin{equation}
  R_{\rm ex} = 1-{\rm H}_2\left(\underline{\delta}_{\rm H eff}\right)\,
\end{equation}
and the cutoff rate bound (\ref{binaryCutOffRateLowerBound}) for rates 
above $R_{\rm ex}$.
\end{theorem}
Proof: The lower bound on the error exponent corresponds to the minimal 
vertical difference between the straight line 
$\underline{\delta}_{\rm H}\log_2\left({1\over c}\right)$ characterizing the 
suboptimal decoding rule (\ref{DH}) and the expurgated IAAME 
(\ref{zeroBelowGV}). It is clear (see Figure~3) that it can
only be attained for 
$\underline{\delta}_{\rm H}\ge\underline{\delta}_{\rm GV}(R)$ since for 
$\underline{\delta}_{\rm H}<\underline{\delta}_{\rm GV}(R)$ the distance is 
infinite. Whenever the function 
$\underline{\delta}_{\rm H}\log_2\left({1\over c}\right) - 
{\rm H}_2(\underline{\delta}_{\rm H}) + 1 - R$ in Theorem~1 attains its 
minimum at 
$\underline{\delta}_{\rm H eff}\le\underline{\delta}_{\rm GV}(R)$, or
equivalently 
$R\le R_{\rm ex}=1-{\rm H}_2\left(\underline{\delta}_{\rm H eff}\right)$, the 
lower bound on the error exponent is given by the expurgated error exponent 
(\ref{GVBH}). Otherwise the lower bound on the error exponent is given by the 
cutoff rate lower bound derived in Theorem~1.
\hfill {} $\Box$

\noindent The representation of the expurgated lower bound on the error exponent
as $\lowE_{\rm ex}(R)=\log_2\left({1\over c}\right) \cdot 
\underline{\delta}_{\rm GV}(R)$ is equivalent to (\ref{implicit}). It can be
recognized as the quantum analog of the 
\emph{Gilbert-Varshamov Bhattacharyya distance} \cite{laz} \cite{omura1} 
\cite{omura2}.

\section{Conclusion}
First we showed that the upper bound on the overall block decoding error 
probability of a binary block code $[N,R]$ used on a binary classical-quantum 
channel $\mathscr{C}$ and decoded by the suboptimal decision rule 
$\mathscr{D}_{\mathcal{H}}$ introduced by Holevo depends on the Hamming 
distances among the code words and the channel parameter $c$. This permited us 
to use the \emph{method of multiplicity enumerators}, derived from 
\cite{laz}, in order to estimate the error exponent and the capacity 
performance of a \emph{specific code family} used on the c-q channel 
$\mathscr{C}$. We showed that the families of binary block codes with 
appropriately \emph{rescaled binomial multiplicity enumerators} attain the 
expurgated and cutoff rate lower bound on the error exponent of binary 
classical-quantum channels with arbitrary signal states over a finite Hilbert 
space. The method of multiplicity enumerators provides a direct and 
conceptually simple method for derivation of these lower bounds. 

For the classical binary symmetric channel is was shown in \cite{bsc} that code
families with rescaled binomial multiplicity enumerators attain the tight
part of the random coding bound and thus the channel capacity. This was 
obtained by an improved upperbounding of the decoding error probability 
assuming the Maximum-likelihood decoding. Unfortunately, this method cannot be 
applied directly to binary classical-quantum channels since the optimal
decision rule for these channels is not known.
\bigskip
\begin{center}{Acknowledgement}\end{center}
P.~Wocjan is supported by Deutsche Forschungsgemeinschaft (DFG) "VIVA".

\end{document}